\documentclass[prd,aps,twocolumn,showpacs]{revtex4}
\newcommand{\be}{\begin{equation}}
\newcommand{\ee}{\end{equation}}
\newcommand{\rf}[1]{(\ref{eq:#1})}
\newcommand{\X}{\bar{\cal{X}}}
\newcommand{\Z}{{\cal Z}}

\input{psfig.tex}
\begin{document}

\title{On the Geometry of Dark Energy}
\author{M. D. Maia}\email{maia@unb.br}
\affiliation{Instituto  de F\'\i sica, Universidade de Bras\'\i
lia,  70919-970, Bras\'\i lia, DF, Brasil}
\author{E. M. Monte}\email{edmundo@fisica.ufpb.br}
\affiliation{Departamento de F\'\i sica, Universidade Federal da
Para\'\i ba, 58059-970, Jo\~ao Pessoa, PB, Brasil}
\author{J. M. F. Maia}\email{jmaia@cnpq.br}
\affiliation{Conselho Nacional de Desenvolvimento Cient\'{\i}fico
e Tecnol\'ogico, SEPN 509, BL A, 70750-901, Bras\'{\i}lia - DF,
Brasil}
\author{J. S. Alcaniz}\email{alcaniz@on.br}
\affiliation{Observat\'orio Nacional,
R. Gal.  Jos\'e Cristino 77, 20921-400, Rio de Janeiro, R.J.  Brasil}

\begin{abstract}
Experimental evidence  suggests that we live in a
spatially flat, accelerating universe composed of roughly one-third of
matter (baryonic + dark) and two-thirds of a negative-pressure
dark component, generically called dark energy. The presence of
such energy not only explains the observed accelerating
expansion of the Universe but also provides the remaining piece of information connecting the inflationary flatness prediction with astronomical observations. However, despite of its  good observational indications,  the nature  of the  dark energy  still remains   an open question.  In this paper we 
explore  a  geometrical explanation for such a component  within the context of  brane-world theory  without mirror symmetry,  leading to  a geometrical interpretation   for  dark  energy  as  warp  in the universe  given  by  the extrinsic  curvature. In 
particular, we study the phenomenological implications  of the extrinsic
curvature of a Friedman-Robertson-Walker universe in a five-dimensional  constant curvature  bulk, with  signatures (4,1)  or  (3,2),
as compared with  the X-matter (XCDM) model. From  the analysis of the  geometrically modified Friedman's  equations,  the deceleration parameter  and the Weak Energy  Condition,  we find a  consistent   agreement  with the presently known  observational data on  inflation for  the deSitter bulk,  but  not  for  the  anti-deSitter  case.

 \end{abstract}
\vspace{3mm} \pacs{04.50+h;98.80cq} \maketitle

\section{Introduction}

        The possibility of an accelerating universe,  as indicated  by  measurements of  SNe Ia, has led  to  one of the most important debates    of modern cosmology, which involves  the conception of a late  time dominant ``dark energy" component with negative pressure  \cite{Perlmutter}. The nature of such dark energy constitutes an open and tantalizing question connecting
cosmology and particle physics. Currently, we do not have a
complete scheme capable of explaining such phenomenon or why it is
happening now. The simplest and most appealing proposal considers
a  relic cosmological constant $\lambda$. However, a reasonable
explanation for the large difference between astrophysical
estimates of this  constant $\lambda/8\pi\,G \approx 10^{-47}
\mbox{Gev}^{4}$ and theoretical estimates for the average vacuum
energy density $\rho_{v} \approx 10^{71} \mbox{Gev}^{4}$ is
unknown,  except through an extreme fine-tuning of  118 orders of
magnitude between  these values \cite{Weinberg,Straumann1}.

Other  more elaborate explanations  for  dark energy have  been proposed.  One  example is  the so
called ``quintessence" model featuring a slowly decaying scalar
field associated with a phenomenological potential with energy scales
of the order of the present day Hubble
constant $\sim 10^{-42}$GeV \cite{Caldwell}. Yet,   it seems
difficult to reconcile such small repulsive  force  with any
attempt to solve the hierarchy problem for fundamental
interactions \cite{Carroll}.

A   phenomenological  explanation based on  current observational data is  given  by the ``x-matter"  or  XCDM model  which is   associated
with  an exotic  fluid characterized by  an equation of  state
 like $p_x = \omega_{x} \rho_x$,
where the  parameter  $\omega_x$ can  be  a  constant or, more generally a function of the time \cite{XCDM}.  The presence of
such a fluid  is  consistent with the observed acceleration rate,
without  conflicting with  the abundance of light elements
resulting from the  big-bang nucleosynthesis
\cite{Carroll}. Although interesting from the
phenomenological point of  view, the XCDM model  lacks  an
explanation from first principles.  

 On the other hand,  we have  witnessed a growing interest in the cosmological implications of  brane-world theory. Generally speaking,   this is  a  gravitational theory defined in a higher-dimensional bulk space  whose  geometry defined by the  the  Einstein-Hilbert principle.
 In such scenario,  standard gauge interactions remain confined to the
four-dimensional space-time (the brane-world  generated by a  3-brane) embedded in the higher
dimensional bulk, but gravitons are free  to probe  the extra  dimensions at  the TeV  scale  of  energies \cite{ADD,RS} (  see \cite{Royreview,LangloisReview} for     recent reviews on brane-world gravity). If such expectations are confirmed, the impact
of  strong and  quantized gravity at the same energy level of the
standard gauge interactions will be considerable. Not only it
eliminates the hierarchical  obstacle for a consistent unification
program but it also suggests a possible laboratory and  cosmic  ray generation   of  branes and   short lived  black holes  by  collision  phenomenology  \cite{Olinto,DimopoulosLandsberg,Cheung,CavagliaRoy}.  A review of  high energy  brane-world  phenomenology can be found in \cite{Lorenzana}.
 
Brane-world theory originated  from    M-theory,  specially  in connection with
 the derivation of  the Horava-Witten   heterotic   $ E_8 \times  E_8$  string theory in the  space $AdS_5 \times  S^5$, through the   compactification of one extra dimension on the  orbifold   $S^1/Z_2$,    using  the  $Z_2$ (or mirror) symmetry on the  circle  $S^1$.   The presence of the   five dimensional  anti-deSitter  $AdS_5$  space  is  mainly motivated    by the  prospects of the  AdS/CFT   correspondence between   the superconformal    Yang-Mills theory in four  dimensions  and  the  anti-deSitter  gravity in five dimensions.

The  same  $Z_2$  symmetry  has  been  used as an  argument to implement brane-world  cosmology  in the  $AdS_5$ bulk,  specially in the popular  Randall-Sundrum model II, 
where  that symmetry is applied across  a  background  brane-world   taken  as   a  boundary embedded in that   bulk.
In this case,  the  
the   extrinsic  curvature  of the  background boundary is completely  determined   by   the  confined  matter energy-momentum tensor, through an algebraic relation  known  as the  Israel-Darmois-Lanczos condition (IDL for  short).

As it happens, when applied  to a  homogeneous and isotropic  cosmology defined  in  the   $AdS_5$ bulk, the  IDL  condition  leads  to  a modification 
 of  Friedman's  equations,  which includes 
 a term   proportional  to the  square of the   energy  density of the  confined perfect fluid of the  universe.   The presence of such quadratic density  was initially welcome  as  a  possible solution to the  accelerated  expansion of the universe. However, soon it was  seen  to be incompatible  with the  big-bang nucleosynthesis, requiring  aditional  fixes \cite{Binetruy1,Binetruy2,Cline}.  More
 recently it  has been shown that high energy  inflationary regimes are  also  constrained by the presence of the  same  quadratic term,
as   compared  with  the  recent data    from  the  SDSS/2dF/WMAP experiments
\cite{TsijukawaLiddle,MaiaCR,Bratt,TsijukawaRoy}.   
It has been also argued that  gravitational  waves generated  by the  bulk geometry may  produce  vector perturbations on the brane-world, whose 
 modes disagree with the data from the same  experiments \cite{RingevalDurrer}.

These  observational constraintes have  suggested  that  brane-world  cosmology using the   $Z_2$  symmetry  and/or   the IDL  condition  should be somehow modified.  For  example,  by adding  a  Gauss-Bonnet  term  to the five dimensional  action,  while still keeping  the  $Z_2$  symmetry \cite{TsijukawaRoy}. Another explored possibility is to  remove  that symmetry  and the IDL  condition  altogether  \cite{DGP,DvaliTurner,MaiaFR}. In a  different approach to the problem, the  $Z_2$  symmetry is   broken   but  some form of   junction condition (including the IDL)  is  maintained \cite{Davis,Gergely,Bowcock,CarterUzan}.
This  has  evolved  to  a more general idea, where  the extrinsic  curvature  should be
 governed by a  dynamical  equation, rather  than just  being  specified at a  background  brane-world \cite{Deruelle,BattyeCarter,BattyeMennim}.

Therefore, the application of    the IDL condition on the brane-world cosmology either  with  the $Z_2$ symmetry or  not,  has  led to  an  extensive and  still ongoing debate, involving  some unresolved issues.
 One of  these  is related to the fact  that  the  IDL  condition expresses the  extrinsic  curvature  in terms of  the confined matter, producing  the  mentioned inflationary constraint. So,  we  may well  ask if  this is  a problem  of the IDL  condition  itself, or if  is  it  a problem inherent to the  extrinsic  curvature and its embedding properties.   If the  IDL  condition is  dropped,  do  we improve the agreement  with the  inflationary data?  If  so,  
 can  we  infer from this data an alternative, perhaps  dynamical,   condition on the extrinsic curvature?   Is the IDL condition an independent postulate? Finally, is the  $Z_2$ symmetry  compatible  with  the  embedding  requirement of the brane-world  structure?

The purpose of this  note is to   investigate   the phenomenology  of the  dark energy  hypothesis   in the   brane-world  context, without  using the  $Z_2$  symmetry,  or without postulating any
junction  condition  separately, at least for the time being. In this case, the  brane-world dynamics  follows  essentially   
from  three basic postulates: the  Einstein-Hilbert 
principle applied to the bulk geometry,
the confinement  hypothesis and the  probing  of the extra dimensions  by  the gravitational field.

 Under these   conditions,  Friedman's  equation  is  modified  by a geometrical term which  is defined by  the  extrinsic  curvature \cite{MaiaGB,MaiaAC}.  In order to  evaluate the compatibility of the resulting  cosmology with  the observations,  we  make an   analogy with the   phenomenological XCDM dark energy model. Based on the analysis of the  deceletarion parameter, we  find that   the  expansion of the  universe  described by  geometrically  modified  Friedman's  equation   can  match   today's  observable data.  We  also find  that   when the  inflation driving energy   is  positive, then  the  universe  expands  in  a  bulk  with  signature $(4,1)$, compatible  with the de Sitter cosmology.
  The more general situations  where  the  
 extrinsic  curvature is  governed by  a  dynamic  condition is examined in a 
  subsequent paper.
 
 As  shown in  the Appendix  A,  the covariant equations  of  motion  are  derived  from the Einstein-Hilbert  action, in  accordance with   the embedding equations,  in the  most general  case.  
The  Appendix  B is specific  to 
 to five dimensions, where  we review 
 the derivation  of the IDL condition,  showing  how  the    $Z_2$ symmetry   specifies  the 
value of the extrinsic  curvature  out of  
Lanczos jump   condition. We  also discuss the    limitations imposed by   $Z_2$ symmetry on the differentiable embedding of the brane-world in a  constant curvature bulk.

\section{The FRW Brane } 

Based on   very general  theorems on differentiable manifold embeddings  \cite{Nash,Greene}, a   five dimensional  bulk with  constant  curvature would have  limited 
degrees  of freedom.  However, in the particular case  of  the Friedman-Robertson-Walker (FRW)  universe  seen   as   a  brane-world, a five-dimensional  bulk  with constant curvature is    sufficient  as it does  not require any  additional  conditions.   The  constant  curvature bulk is    characterized  by the Riemann
tensor$^{[1]}$ \footnotetext[1]{A curly   ${\cal R}$  denotes bulk
curvatures while  a straight $R$  denotes   brane-world
curvatures. Capital  Latin indices  refer to the bulk dimensions.
Small case Latin  indices refer to the extra dimensions and  all
Greek indices refer to the brane. The semicolon    denotes the
covariant derivative  with respect to $g_{\alpha\beta}$. For
generality  we denote ${\cal G}=|\mbox{det}( {\cal G}_{AB})|$ }
\[
{\cal R}_{ABCD} =  K_{*} ({\cal G}_{AC}{\cal G}_{BD}-{\cal
G}_{AD}{\cal G}_{BC}),  \label{eq:CC}
\]
where  ${\cal G}_{AB}$ denotes the  bulk metric components in
arbitrary coordinates and where  $K_{*}$ is either  zero (for a
flat bulk),  or  it can be proportional  to  a positive or  negative   a bulk cosmological  constant, respectively corresponding 
to the  two possible
signatures:  $(4,1)$  for the  deSitter $dS_5$ bulk and   and  $(3,2)$ for the  anti-deSitter bulk  $AdS_5$.
 Accordingly,  we take   in  the  embedding  equations  \rf{X} in the Appendix  A, $g^{55}=\varepsilon =\pm 1$.  In any of these 
 cases,  the     integrability conditions for the embedding, equations   \rf{Gauss}-\rf{Ricci}  become simply:
\begin{eqnarray}
&&\hspace{-7mm}R_{\alpha\beta\gamma\delta} =\!\!
\frac{1}{\varepsilon} (k_{\alpha\gamma}k_{\beta\delta}\!-\!
k_{\alpha\delta}k_{\beta\gamma})\!\! +\!\!
K_{*} (g_{\alpha\gamma}g_{\beta\delta}-g_{\alpha\delta}g_{\beta\gamma})\label{eq:G1}\\
&&\hspace{-7mm}k_{\alpha[\beta;\gamma]} = 0, \label{eq:C1}
\end{eqnarray}
where  the  sign  of  the last term in  \rf{G1}  depends on the sign of   $K_{*}$.   
The   equations of  motion derived 
in appendix  A can be   adjusted  to the present case,  but it is  just as easy  
to derive  directly from    \rf{G1} and  \rf{C1}. The result is   essentially Einstein's  equations  as  modified by the  presence of  the  extrinsic  curvature. 
(For the  covariant equations in  five dimensions in a more general setting, see
  \cite{Maeda}):
\begin{equation}
\label{eq:Einstein2}R_{\mu\nu}-\frac{1}{2}Rg_{\mu\nu}+\lambda
g_{\mu\nu} = -8\pi G T_{\mu\nu} + {Q}_{\mu\nu},
\end{equation}
where   we have denoted  by $\lambda$ the effective   cosmological constant in four dimensions, including the confined vacuum energy.
  The last term in \rf{Einstein2} 
derived from  \rf{Qmunu} in  the appendix  A is 
   completely geometrical:
\begin{equation}
Q_{\mu\nu}\!\!  =  \!\!\frac{1}{\varepsilon}( k^{\rho}{}_{\mu }k_{\rho\nu
}-h k_{\mu\nu}\!\!-\!\!\frac{1}{2}(K^{2}-h^{2})g_{\mu\nu}),\label{eq:Qij}
\end{equation}
Here we  have  denoted $h=  g^{\mu\nu}k_{\mu\nu}$ and
$K^{2}=k^{\mu\nu}k_{\mu\nu}$.

For the  purpose of  the  embedding of the FRW  universe 
in a  five-dimensional bulk with maximal symmetry
 it is  convenient to parametrize the  FRW metric  as  \cite{Rosen}
\[
dS^{2}=g_{\alpha\beta}dx^{\alpha}dx^{\beta}=-dt^{2}
+a^{2}[dr^{2}+f(r)(d\theta^{2} +sen^{2}\theta d\varphi^{2})]
\]
where  $f(r)=\sin r, r, \sinh r$   corresponding to the spatial
curvature  $k=1,0,-1$ respectively  and  where the confined  source  is the perfect fluid given  in co-moving coordinates  by
 \begin{equation}
T_{\alpha\beta}=(p+\rho)U_{\alpha}U_{\beta}
+pg_{\alpha\beta},\;\;U_{\alpha}=\delta_{\alpha}^{4}. \label{eq:T}
\end{equation}

 Using York's  relation (eqn. \rf{york} in  Appendix  A)  it follows that in the FRW space-time
$k_{\alpha\beta}$ is diagonal.
After separating
the spatial components  we find that Codazzi's  equations (\ref{eq:C1}) reduce to  (here $\mu, \nu, \rho, \sigma = 1..3$).
\begin{eqnarray*}
 k_{\mu\nu,\rho}\!\!&- & k_{\nu\sigma}\Gamma^{\sigma}_{\mu\rho}= k_{\mu\rho,\nu}
-k_{\rho\sigma}\Gamma^{\sigma}_{\mu\nu}, \\
{k}_{\mu\nu,4}\!\!&- &k_{\mu\nu}\frac{\dot a}{a}
=-a\dot{a}(\delta^{1}_{\mu}\delta^{1}_{\nu}
 \!+\!f^{2}\delta^{2}_{\mu}\delta^{2}_{\nu} \!+\! f^{2}\sin^{2}\theta \;
 \delta^{3}_{\mu}\delta^{3}_{\nu})k_{44}.
\end{eqnarray*}
where  $a(t)$ is the scale factor and the  dot  means derivative with respect to $t$.
 The first equation   gives
$k_{11,\nu} =0$ for  $\nu\neq 1$,  so that $k_{11}$ does not
depend on  the  spatial coordinates. Denoting  $k_{11}=b(t)$,  the
second  equations give \cite{MaiaMI}
\begin{equation}
 k_{44}=-\frac{1}{\dot{a}}
\frac{d}{dt}\left(\frac{b}{a}\right)\label{eq:kij}.
\end{equation}
Repeating the same procedure for  $\mu, \nu = 2,3$  we  obtain  $k_{22}=b(t) f^{2}$ and
$k_{33}=  b(t)f^{2}\sin^{2}\theta$.
In short, 
  \be
   k_{\mu\nu}=\frac{b}{a^{2}}g_{\mu\nu},\;\;\; \mu,
\nu,  = 1\ldots 3 \label{eq:kab}
\ee 
Thus,  \rf{kij} and  \rf{kab} represent  the general solution of  \rf{C1} for the FRW metric  in  a  5-dimensional  constant curvature bulk.  
Notice that as   a  consequence of the homogeneity of
\rf{C1}, the function $b(t)=k_{11}$ representing the extrinsic  curvature component along the  radial
direction (r),  remains  arbitrary.   Denoting  $B=\dot{b}/b$ e  $H=\dot{a}/a$,  we find from
 \rf{Qij}  that 
\begin{equation}
K^{2}= \frac{b^{2}}{a^{4}} ( \frac{B^{2}}{H^{2}}-2\frac{B}{H}+4),
\;\; h=\frac{b}{a^{2}}(\frac{B}{H}+2) \label{eq:K}
\end{equation}
\begin{eqnarray}
Q_{\mu\nu}&=&
\frac{1}{\varepsilon}\frac{b^{2}}{a^{4}}\left(2\frac{B}{H}-1\right)g_{\mu\nu},\;\;\mu,\nu =1..3   \label{eq:Qmunu5}\\
 Q_{44} &=& -\frac{1}{\varepsilon}\frac{3b^{2}}{a^{4}} \label{eq:Qab}\\
Q&=& g^{\alpha\beta}
Q_{\alpha\beta}=\frac{1}{\varepsilon}\frac{6b^{2}}{a^{4}}\frac{B}{H}
\label{eq:Q2}
\end{eqnarray}
Replacing  these  in   \rf{Einstein2} and
separating   the space and  time  components it follows that
\begin{eqnarray*}
&&\frac{\ddot{a}}{a} +2\frac{\dot{a}^{2}}{a^{2}} +2\frac{k}{a^{2}}
=4\pi G (\rho\!-\! p) +\frac{\lambda}{2}+\frac{1}{\varepsilon}
\frac{b^{2}}{a^{4}}
\frac{1}{\dot{a}b}\frac{d}{dt}(ab)\\
&&\frac{\ddot{a}}{a} =-\frac{4\pi G}{3}(\rho + 3p)
+\frac{\lambda}{6} + \frac{1}{\varepsilon}\frac{b^{2}}{a^{4}}
\frac{1}{\dot{a}b}a^{2}\frac{d}{dt}(\frac{b}{a})
\end{eqnarray*}
and finally, after eliminating $\ddot a$,  we obtain  the
modified Friedman's equation for the  FRW brane-world
\begin{equation}
(\frac{\dot{a}}{a})^2+\frac{k}{a^{2}}=\frac{8\pi G}{3}\rho
+\frac{\lambda}{3}+ \frac{1}{\varepsilon}\frac{b^2}{a^4}
\label{eq:Friedmann}
\end{equation}
where we  see that  the   correction term with respect  to the  standard  Friedman's  equation is given by   the   component of the extrinsic curvature$^{[2]}$ 
\footnotetext[2]{
Just for  comparison purposes, it is  illustrative  to  see 
how the     $\rho^{2}$ term may emerges from \rf{Friedmann}  when    the
IDL  condition (equation  \rf{israel} in  appendix B)  is   postulated. For  the perfect fluid with energy density  $\rho$, that 
condition  gives  $k_{11}=b(t)
=-\alpha_{*}^{2}\rho a^{2}$. Replacing this in   \rf{Friedmann} we obtain 
\[
(\frac{\dot a}{a})^2+\frac{k}{a^{2}} =\frac{8\pi G}{3}\rho +
\frac{\lambda}{3} + \frac{1}{\varepsilon} \alpha_{*}^{4} \rho^{2}
\]
producing the    $\rho^{2}$ dependent cosmology \cite{MaiaFR}. }.  Notice also the  presence of
$\varepsilon$ which marks  the effects  of the bulk signature on
the expansion of the universe.

\section{Dark Energy as  Geometry}

The additional degrees of freedom  offered  by  brane-world gravity admits   a  wide range  of possibilities for  dark energy, beyond the  $\Lambda$CDM model  \cite{Sahni}. Here  we explore the  fact that
$Q_{\alpha\beta}$  is  independently  conserved, suggesting   an  analogy with  the energy-momentum
tensor of   an uncoupled  non-conventional  energy source. In this  analogy  we take the
 XCDM model as  a practical example,  denoting    the "geometric pressure"  associated with  the  extrinsic curvature by 
$p_{extr}$ (the
 suffix  $extr$  stands for ``extrinsic") and the "geometric energy density" by  
$\rho_{extr}$.  The  corresponding geometric energy-momentum is identified   to  $Q_{\mu\nu}$
 as
\begin{equation}
Q_{\mu\nu}\!\! \stackrel{def}{=}\!\! -\frac{1}{8\pi G}\left[(p_{extr} + \rho_{extr})
U_{\mu}U_{\nu} +p_{extr}g_{\mu\nu}\right]  \label{eq:exfluid}
\end{equation}
where $U_{\alpha}=\delta^{4}_{\alpha}$.
Comparing   with the previous components \rf{Qmunu5}-\rf{Q2}
  we obtain
\begin{eqnarray}
\hspace{-3mm}p_{extr} \!\!  = \!\!  -\frac{1}{8\pi
G\varepsilon}\frac{b^{2}}{a^{4}}\left( 2\frac{B}{H}-1\right), \;\;
\;\; \rho_{extr}\! =\! \frac{3}{8\pi
G\varepsilon}\frac{b^{2}}{a^{4}}
\end{eqnarray}
Notice  the    dependence of these terms on  the bulk
signature  $\varepsilon$. The  sign  $(-)$ in  \rf{exfluid}
was chosen  in  accordance with the weak energy condition corresponding to the positive energy $\rho_{extr}>0$ and  negative pressure $p_{extr}<0$  with  $\varepsilon =1$.

Like  the   XCDM model, the  geometric
``dark energy  fluid" can be implemented  by a state-like equation
\begin{equation}
p_{extr} = \omega_{extr} \rho_{extr}  \label{eq:state}
\end{equation}
where  $\omega_{extr}$ may be  a function
of time. After replacing the expressions of $B$ and
$H$,  we obtain the following equation for  $b(t)$ \be
\frac{\dot{b}}{b}=
\frac{1}{2}(1-3\omega_{extr})\frac{\dot{a}}{a}\label{eq:eforb}.
\ee 
We cannot  readily solve this  equation because  $\omega_{extr}$  is not known.  However,  a   simple and  useful example is given  when  $\omega_{extr}=\omega^{0}_{extr}$=constant. In  this case, the  general solution of   \rf{eforb} is  very  simple:
 \be 
 b(t) =
b_{0}(\frac{a}{a_{0}})^{\frac{1}{2}(1-3\omega^{0}_{extr})} \label{eq:b}
\ee
where $a_{0}$   is  the present  value  of  the  expansion
parameter  and  $b_{0}$ is an integration constant representing the current  warp of the universe.  Clearly  it must not vanish,
otherwise all extrinsic  curvature  components
would also vanish, and the brane-world would behave  just a trivial plane.

Replacing s \rf{b} in   \rf{Friedmann}, now expressed  in    terms of the  redshift $z =
{a_o}/a - 1$  and of the observable parameters  $\Omega$'s, we obtain
\begin{eqnarray}
(\frac{\dot{a}}{a})^{2} & = & H_o^{2}[\Omega_m(1 + z)^{3} + \Omega_{\lambda} + \\
\nonumber & & \quad         \quad   \quad         \quad +
\frac{1}{\varepsilon}\Omega_{extr} (1 + z)^{3(1 +
\omega^{0}_{extr})}  + \Omega_k(1 + z)^{2}],
\end{eqnarray}
where  $H_o$ is the present value of the Hubble parameter,
$\Omega_m$, $\Omega_{\lambda}$ and $\Omega_k$ are, respectively,
the  confined  matter, cosmological constant  and the spatial
curvature relative density parameters  and where
we have denoted the  relative  density parameter associated with
the  geometrical  dark energy by \be \Omega_{extr} =
\frac{{b_{o}}^{2}}{H_o^{2}a_o^{4}}, \ee

Notice that  equation \rf{eforb} is identical to the corresponding equation in  XCDM \cite{XCDM}, except  that  here $b(t)$ has the geometrical meaning of  the
radial component of the extrinsic  curvature  $k_{11}$. 

If  we had taken   the  bulk signature  to be  $(3,2)$ (or, $\varepsilon
=-1$), equation \rf{eforb} would  represent    a fluid   with  negative energy and  positive pressure, producing an  unexpected
behavior  of the expansion of the universe. In order to better
visualize this difference, figures 1a and 1b show the behavior of
the deceleration parameter $q(z) = - \ddot{a}a/\dot{a}^{2}$ as a
function of redshift for selected values of $\Omega_m$ and
$\Omega_{extr}$. Figure 1a shows the plane $q(z) - z$ for the
signature $\varepsilon = +1$. As it is well known, the acceleration redshift for such models happens
around $z_a \simeq 0.7$, which seems to be in good agreement with
observational data \cite{triess}.  However, the case $\varepsilon
= - 1$ in Figure 1b presents an opposite behavior, with the
deceleration parameter becoming more positive at redshifts of the
order of $z \simeq 1$. Therefore, in the light of this simple
qualitative analysis and having in mind the recent supernovae
(SNe) results \cite{Perlmutter}, it is possible to exclude the
bulk signature  $(3,2)$ for  an  acceleration  driven by  a positive  $\rho_{extr}$.
\begin{figure}
\vspace{.2in}
\centerline{\psfig{figure=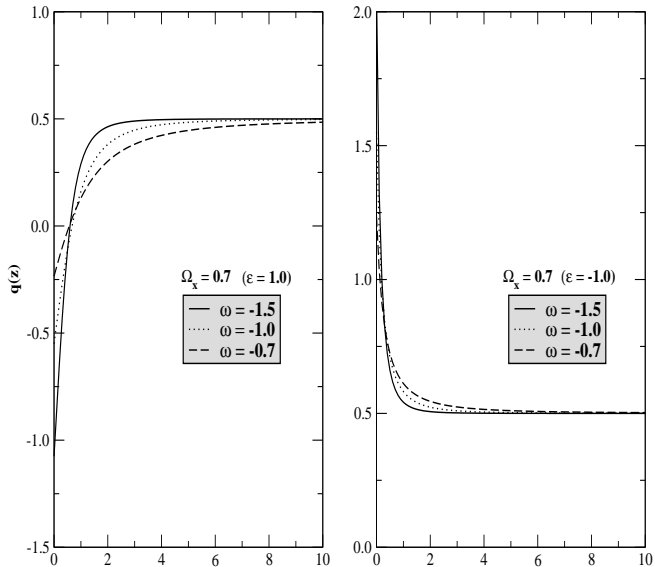,width=3.5truein,height=3.3truein,angle=-90}
\hskip 0.1in} \caption{Deceleration parameter as a function of
redshift for a fixed value of $\Omega_x = 0.7$. In both Panels the
horizontal line labeled decelerating/accelerating ($q_o = 0$)
divides models with a decelerating or accelerating expansion at a
given redshift. Note that while the signature  $\varepsilon = 1$
gives a speed up  scenario at $z_a \simeq 1$, the signature
$\varepsilon = - 1$ implies a  slow down scenario}
\end{figure}

The use of  the bulk signature $(4,1)$  associated with $\rho_{extr}>0$  allows  us  to  use the wealth of  available data  from  the recent  measurements 
 to determine  limits  on the values of  $\omega^{0}_{extr}$ in our   geometric model.  For example, from the current SNe Ia data (the so-called \emph{gold} set of 
157 events) one finds $\omega^{0}_{extr} < -0.5$ at 95\% confidence level 
(c.l.) for $\Omega_{\lambda} = \Omega_k = 0$, regardless of the value of the 
matter density parameter \cite{riess}.  When combined with Cosmic Microwave 
Background (CMB) and Large-Scale Structure (LSS) observations, the same SNe Ia 
data provide $\omega = -1.02^{+0.13}_{0.19}$ (and $\omega < -0.76$ at 95\% 
c.l.) \cite{garn}.  These limits  agree  with the constraints obtained from a wide variety 
of different phenomena, using the ``cosmic concordance method" \cite{wang}. In 
this case, the combined maximum likelihood analysis suggests $\omega^{0}_
{extr} \leq -0.6$, (which incidently also rules out an unknown component like 
topological defects of  dimension  $n$, such as domain walls and cosmic 
strings, for which we would have $\omega^{0}_{extr} = -\frac{n}{3}$).  Other  
methods have  also  contributed  to  the collection of data. For example,  
gravitational lens statistics based on the final Cosmic Lens All Sky Survey 
data suggests that $\omega^{0}_{extr} < -0.55^{+0.18}_{-0.11}$ at 68\% c.l. 
\cite{chae,abha}. Similarly, distance estimates of galaxy clusters from 
interferometric measurements of the Sunyaev-Zel´dovich effect and X-ray 
observations along with SNe Ia and CMB data requires $\omega = -1.2^{+0.11}_{-
0.18}$ \cite{schu,alcaniz2003,alcaniz2003b}. We may also use the  measurements 
of the angular size of high-redshift sources, suggesting   that we could take  
$-1 \leq \omega^{0}_{extr} \leq -0.5$ \cite{jsa}, whereas the use of SNe data 
and measurements of the position of the acoustic peaks in the CMB spectrum, 
suggest $-1\leq \omega^{0}_{extr} \leq-0.93$ at 2$\sigma$ \cite{cora}. We, 
therefore, conclude that  in  contrast  with the five-dimensional brane-world  
models with the $\rho^{2}$  term in Friedman's equation  which  face  
experimental  constraints, the present geometrical  model, at least in the 
simple case where $\omega^{0}_{extr}$  constant, is consistent with the latest 
experimental observations,  within the  limits  imposed by the weak energy  condition,  in the  deSitter bulk.

\section{Summary}

We  have  provided  ample  experimental evidence  in support  to  the hypothesis  that  dark energy  can  be   a  consequence of the extrinsic  curvature  in brane-world  cosmology.
   For that purpose,  we have taken  the   FRW universe,  seen  as  a  brane-world embedded in a five dimensional bulk of  constant  curvature,   with undefined  signature and  the  without  the  $Z_2$  symmetry or any  form of   junction  condition. The  indefined  signature  has been motivated mostly  by  
the fact that, except from the theoretical  arguments in the application of  the  AdS/CFT in  string  theory, there is   no  experimental  argument in favor  of the  anti-deSitter signature $(3,2)$
over  the  deSitter  signature  $(4,1)$.

Under these  conditions,     Friedman's  equation   depends  only on the signature  of  the bulk  and on  the  extrinsic  curvature  as  a possible  driver  for  inflation  and.  In order to evaluate the compatibility of such geometrical model  with the present   experimental data,  we  have established  a  correspondence  with  the   phenomenological      XCDM dark energy  model.

In  the simple example where the   factor $\omega_{extr}$  in the state  equation   \rf{state} is constant,  we  found  that when  the  energy density of the geometric fluid  is   positive and the pressure is negative,       the $AdS_5$ bulk  with  signature  $(3,2)$  is not   compatible with  the expected value of the deceleration parameter for  the  redshift $z\approx 1$,  favoring the  de  Sitter case with signature $(4,1)$.  However,  if we had  taken the  expansion energy to be negative, with positive pressure,  then  the universe would  expand  in  the bulk  with signature
$(3,2)$.

This  example  suggests  not only  that  the extrinsic  curvature  can be   the responsible  for the  accelerated  expansion,  but also  
that  in the more general case where  $\omega_{extr}$ is  not  constant  it  mus be  
dynamic,  much in  the sense  proposed  in the literature. The fact  that   the  extrinsic  curvature
is  a symmetric rank-two  tensor  suggests  
that  the  required  dynamical  equation 
  should be   non-linear, in fact  an Einstein-like  equation \cite{Gupta}.   Work on such   ultimate dynamical "junction"  condition  is  still in progress,

\phantom{x}\hrulefill

\begin{center}
{\textbf{Appendix  A:\\  Equations of Motion}}
\end{center}
With  a few  exceptions mostly in the five-dimensional cases,  the use of  differentiable  properties  of the brane-world embedding  have  been  largely neglected.  Nonetheless, the probing  of the   extra  dimensions by gravitons  
means that  the classical geometry   of the brane-world   should be  subjected to   continuous  deformations  (or  perturbations) in response to the   Einstein-Hilbert dynamics of the bulk.  Such  deformable embeddings   have been  studied  by  Nash and Greene, concluding that 
the dimension and  signature of the  bulk  is not a matter of choice,  but they depend  on the  the  regularity  of the  embedding  functions  \cite{Nash,Greene}.  Therefore,  restriction  of the bulk  to  the  $AdS_5$
geometry requires  the use  of  some  additional  conditions,  like imposing boundary  rigidity, and the application of the  $Z_2$  symmetry. However,  it  not clear  that  the  full dynamics of  the brane-world will  be  compatible  with  such  limited  embedding, except perhaps  in   some  specific cases, as  the FRW  example in the  main text.  The  embedding  is,  of  course, a  fundamental issue  in differential  geometry  which  has been frequently  applied  to  general  relativity. A  comprehensive reference list on space-time embedding properties  can be found in  \cite{Pavsic}.
In  the  following we  give  a short  summary  on the generation of  
the differentiable  embedding  by perturbations  of  a given background geometry.

 Denoting   by  $\bar{g}_{\alpha\beta}$  the  metric  of  a given four-dimensional   manifold  $\bar{V}_{4}$ (the background)  and  by  ${\cal G}_{AB}$  the metric of  the Riemannian bulk $V_{D}$,  the isometric  embedding  of  $\bar{V}_{4}$   is  given  by   a  map   ${\cal X}: \bar{V}_{4}\rightarrow  V_D$,  with $D=4+N$ components  $\bar{\cal  X}^{A}$ 
 such that 
 \be
\!\!{\X}^{A}_{,\mu}{\X}^{B}_{,\nu}{\cal G}_{AB}
=\bar{g}_{\mu\nu},\;
{\X}^{A}_{,\mu}\bar{\eta}^{B}_{b}{\cal G}_{AB}=0,\;
\bar{\eta}^{A}_{a}\bar{\eta}^{B}_{b}{\cal
G}_{AB}=\bar{g}_{ab}  \label{eq:X}\!\!\!\!
\ee
where $\bar{\eta}^{A}_{a}$  are the  components of the $N$
independent normal vectors to $\bar{V}_{4}$.  According to Nash,   we may
continuously deform  $\bar{V}_{4}$  along a
normal  direction  in the bulk, to obtain
another submanifold of the same  bulk,   provided the
 embedding functions  remain regular (in a
generalized sense).  
 Denoting  a  generic orthogonal  direction by ${\eta} =y^{a}\bar{\eta}_{a}$, the  deformation  (in fact a perturbation)  of the embedding  can be  expressed  as
\[
\Z^{A}  =\X^{A}  +(\pounds_{\eta}\X)^{A}, \;\;
\eta^{A}=\bar{\eta}^{A} + (\pounds_{{\eta}}\bar{\eta})^{A} =
\bar{\eta}^{A}
\]
The  components  ${\cal Z}^{A}$  and  the  normal
$\eta_{a}$ must  satisfy   embedding equations similar to  \rf{X} (now  dependent on  $y^a$):
\begin{equation}
\phantom{x}\hspace{-0.5cm}\!{\cal Z}^{A}_{,\mu}{\cal
Z}^{B}_{,\nu}{\cal G}_{AB}\!\! =\!\!g_{\mu\nu},\; {\cal
Z}^{A}_{,\mu}{\eta}^{B}_{a}{\cal G}_{AB}\!\!=\!g_{\mu a},\;\; {\eta}^{A}_{a}{\eta}^{B}_{b}{\cal G}_{AB}\!\!=\!\!g_{ab}
\!\!\!\! \label{eq:Z}
\end{equation}
From  these equations   it follows that
\begin{equation}
g^{\mu\nu}Z^{A}_{,\mu}Z^{B}_{,\nu} = {\cal G}^{AB}
-g^{ab}\eta^{A}_{a}\eta^{B}_{b}\label{eq:invert}
\end{equation}
and also  the  components of the   perturbed  geometry
 \begin{eqnarray}
g_{\mu\nu}(x,y) & = &
{\mathcal{Z}}_{,\mu}^{A}{\mathcal{Z}}_{,\nu}^{B}{\mathcal{G}}_{\!
AB}=\bar{g}_{\mu\nu}\!\!- 2y^{a}\bar{k}_{\mu\nu a}  + \nonumber \\
 &&\hspace{-8mm}
y^{a}y^{b}[\bar{g}^{\alpha\beta}\bar{k}_{\mu\alpha a}\bar{k}_{\nu\beta b}+g^{cd}\bar{A}_{\mu
ca}\bar{A}_{\nu db}],\label{eq:gmunu}\\
g_{\mu a}(x,y) & = & 
{\mathcal{Z}}_{,\mu}^{A}{\eta}_{a}^{B}{\mathcal{G}}_{\! AB}=\!\! y^{b}A_{\mu ab},  \label{eq:gmua}\\
g_{ab}(x,y) & = & 
{\eta}_{a}^{A}{\eta}_{b}^{B}{\mathcal{G}}_{\! AB}= \bar{g_{ab}} \label{eq:gab}\\
k_{\mu\nu a}(x,y) & = &
-\eta_{a,\mu}^{A}{\mathcal{Z}}_{,\nu}^{B}{\mathcal{G}}_{AB} = \nonumber\\ 
\hspace{-6mm}&&\phantom{x}\hspace{-10mm}\bar{k}_{\mu\nu
a}\!\!-y^{b}\bar{g}^{\alpha\beta}\bar{k}_{\mu\alpha a}\bar{k}_{\nu\beta
b}-\!\!{g}^{cd}y^{b}\bar{A}_{\mu ca}\bar{A}_{\nu db},\hspace{3mm}\label{eq:kmunua} \\
A_{\mu ab}(x,y) & = &
\eta_{a,\mu}^{A}\eta_{b}^{B}{\mathcal{G}}_{AB}\!\!=\!\!\bar{A}_{\mu ab}(x)\label{eq:A}
\end{eqnarray}
From   \rf{gmunu} and \rf{kmunua} we obtain  York's relation for the extrinsic  curvature (extended to the extra variables
$y^{a}$):
\begin{equation}
k_{\mu\nu a}=-\frac{1}{2}\frac{\partial g_{\mu\nu}}{\partial y^{a}} \label{eq:york}
\end{equation}
so that when the   brane-world  gravitational field  represented  by the  metric  $g_{\mu\nu}$  propagates in the bulk, so  does the   extrinsic  curvature.

The components of the Riemann tensor of the bulk written in the
the embedding vielbein $\{\Z^{A}_{,\alpha},\eta^{A}_{a} \}$, give the Gauss, Codazzi and Ricci  equations,  respectively
\cite{Eisenhart}:
\begin{eqnarray}
&&\phantom{x}\hspace{-1,5cm} R_{\alpha\beta\gamma\delta}\!\!
=\!\! 2g^{mn}k_{\alpha[\gamma m}k_{\delta]\beta n}
\!\!+{\cal R}_{ABCD}{\cal Z}^{A}_{,\alpha}
{\cal Z}^{B}_{,\beta}{\cal Z}^{C}_{,\gamma}{\cal Z}^{D}_{,\delta} \label{eq:Gauss}\\
&&\phantom{x}\hspace{-1,6cm}k_{\alpha[\gamma b; \delta]}\!\! =\!\!
g^{mn}A_{[\gamma mb}k_{\alpha\delta]n } \!\!+{\cal R}_{ABCD} {\cal
Z}^{A}_{,\alpha} \eta^{B}_{b}{\cal Z}^{C}_{,\gamma}{\cal
Z}^{D}_{,\delta} \label{eq:Codazzi}\\
&&\phantom{x}\hspace{-1,8cm}2A_{[\gamma a b ; \delta]}\! =\!\!
- 2g^{mn}A_{[\gamma ma}A_{\delta]n b}
\nonumber \\
&-& g^{mn}k_{[\gamma m a}k_{\delta]nb}\!\! - {\cal
R}_{ABCD}\eta^{A}_{a}\eta^{B}_{b} {\cal Z}^{C}_{,\gamma} {\cal
Z}^{D}_{,\delta} \label{eq:Ricci}
\end{eqnarray}
To proceed,  we now  impose that the bulk geometry is  a  solution of  Einstein's  equations.  Denoting $K^{2}=g^{ab}k^{\mu\nu}{}_{a}k_{\mu\nu b}$,
$h_{a}=g^{\mu\nu}k_{\mu\nu a}$ and $h^{2} =g^{ab }h_{a}h_{b}$ and
using \rf{invert},  we obtain  from the contraction of Gauss'
equations  with  $g^{\mu\nu}$
\begin{eqnarray}
R_{\mu\nu} & = &g^{cd}(g^{\alpha\beta}k_{\mu\alpha c} k_{\nu \beta
d} - h_{c}k_{\mu\nu d})  +{\cal  R}_{AB}Z^{A}_{,\mu}Z^{B}_{,\nu}
\nonumber \\& -& g^{ab}{\cal R}_{ABCD}\eta^{A}_{a}
Z^{B}_{,\mu}Z^{C}_{,\nu}\eta^{D}_{b}\label{eq:RicciT}
\end{eqnarray}
A  further contraction gives the  Ricci scalar
\begin{eqnarray}
R & =& (K^{2} -h^{2})  +{\cal R}  -2g^{ab}{\cal R}_{AB}
\eta^{A}_{a}\eta^{B}_{b}
\nonumber\\
 & + & g^{ad}g^{bc}{\cal R}_{ABCD}\eta^{A}_{a}\eta^{B}_{b}\eta^{C}_{c}\eta^{D}_{d}
\label{eq:RICCI}
\end{eqnarray}
Therefore,  the Einstein-Hilbert action for the bulk geometry in $D$-dimensions can be  written  in terms of the  embedded  geometry as 
\begin{eqnarray}
&&\int{\cal R}\sqrt{\cal{G}}d^{D}v \equiv
\int\left[ R -( K^{2} - h^{2})\right]\sqrt{\cal{G}}d^{D}v\nonumber  +\\
&&\phantom{x}\hspace{-1cm}\!\!\!\!\!\int \left[ 2g^{ab} {\cal
R}_{AB}\eta^{A}_{a}\eta^{B}_{b}\!\!\! - \! g^{ad}g^{bc}{\cal
R}_{ABCD}\eta^{A}_{a}\eta^{B}_{b}
\eta^{C}_{c}\eta^{D}_{d} \right] \sqrt{\cal G}d^{D}v\nonumber\\
&&=\alpha_{*}\int {\cal L}^{*}\sqrt{\cal{G}}d^{D}v \label{eq:EH}
\end{eqnarray}
where in the  right hand  side  we have  included 
the bulk source Lagrangian  ${\cal
L}^{*}$ and  we have  denoted  by $\alpha_{*}$  the D-dimensional fundamental energy scale.

The covariant equations of motion for
a brane-world  in a  D-dimensional bulk  can be derived  by taking the variation of  \rf{EH} with  respect to $g_{\mu\nu}$, $g_{\mu  a}$  and   $ g_{ab}$, noting that  the Lagrangian  depends on these variables through ${\cal Z}^{A}_{,\mu}$. Alternatively, wemay just  calculate  the components  of  Einstein's  equations for the bulk  geometry:
\begin{equation}
{\mathcal{R}}_{AB} -\frac{1}{2}{\mathcal{R}}{\mathcal{G}}_{AB}= \alpha_{*}T_{AB}^{*}
\label{eq:bulkEE2}
\end{equation}
in the  embedding vielbein  $\{ {\cal Z}^A_{,\alpha},  \eta^B_{b}\}$. Denoting the vielbein
 components of the energy-momentum tensor  by 
$T_{\mu\nu}^{*}=T^{*}_{AB}{\cal Z}^{A}_{,\mu}{\cal Z}^{B}_{,\nu}$, $T_{\mu a}^{*}=T^{*}_{AB}{\cal Z}^{A}_{,\mu}\eta^{B}_{a}$ and  $T_{ab}^{*}=T^{*}_{AB}\eta^{A}_{a}\eta^{B}_{b}$,  we obtain  from \rf{bulkEE2} 
\[
g^{ab}{\cal R}_{AB}\eta^{A}_{a}\eta^{B}_{b}  =  \alpha_* ( g^{ab} T^*_{ab} -\frac{N}{N+2}T^*)
\]
From  \rf{RicciT}  and  \rf{RICCI}, the (tangent)
components of \rf{bulkEE2} on ${\cal Z}^{A}_{,\mu}{\cal Z}^{B}_{,\nu}$  give the   equation for  $g_{\mu\nu}$ (sometimes referred to as the \emph{gravi-tensor}  equation) 
\begin{eqnarray}
&&\phantom{x}\hspace{-10mm}R_{\mu\nu}-\frac{1}{2}R
g_{\mu\nu}-Q_{\mu\nu}- (W_{\mu\nu}-\frac{1}{2}W g_{\mu\nu})
\nonumber \\
&&\phantom{x}\hspace{1cm}=  \alpha_* (T^*_{\mu\nu}-\frac{N}{N+2}T^* g_{\mu\nu}-g^{ab}T^*_{ab}),  \label{eq:BE1}
\end{eqnarray}
 where  we have denoted
\begin{eqnarray}
Q_{\mu\nu}\!\! & = & \!\! g^{ab}k^{\rho}{}_{\mu a}k_{\rho\nu
b}-g^{ab}h_{a}k_{\mu\nu b}\!\!-\!\!\frac{1}{2}(K^{2}-h^{2})g_{\mu\nu}
\label{eq:Qmunu}\\
\!\!W_{\mu\nu} & = & \!\!
g^{ad}{\mathcal{R}}_{ABCD}\eta_{a}^{A}{\mathcal{Z}}_{,\mu}^{B}{\mathcal{Z}}_{,\nu}^{C}\eta_{d}^{D} \nonumber\\
 W & = & g^{ad}g^{bc}{\mathcal{R}}_{ABCD}\eta_{a}^{A}\eta_{b}^{B}\eta_{c}^{C}\eta_{d}^{D} \nonumber
\end{eqnarray}

On the other hand,  again from  \rf{bulkEE2},  the trace  of 
Codazzi's  equation  \rf{Codazzi} gives the  \emph{gravi-vector}  equation
\begin{eqnarray}
 &&\phantom{x}\hspace{-8mm}k_{\mu a;\rho}^{\rho}\! -\!h_{a,\mu}   \!+\! A_{\rho c a}k^{\rho \;c}_{\;\mu}\! -\!A_{\mu c  a}h^{c}   
 \!\!  +\!\! 2 W_{\mu a}\\&&\phantom{x}\hspace{8mm} = \alpha_* (T^*_{\mu a}  -\frac{1}{N+2}T^* g_{\mu a})\label{eq:BE2}
\end{eqnarray}
where   
$$  W_{\mu  a}= g^{mn} {\cal R}_{ABCD}\eta^{A}_{a}\eta^{B}_{m}{\cal Z}^{C}_{,\mu}\eta^{D}_{n}   $$
Finally, the \emph{gravi-scalar} equation is obtained from \rf{RICCI}
and \rf{bulkEE2} 
\begin{eqnarray}
&&\phantom{x}\hspace{-10mm}R-K^{2} +h^{2} - W= -2\alpha_*(g^{ab}T^*_{ab} - \frac{N+1}{N+2}T^*)
\label{eq:BE3}
\end{eqnarray}
Equations \rf{BE1}-\rf{BE3} represent the most general  equations of motion of  a brane-world,  compatible  with the
its differentiable  embedding in a  D-dimensional bulk defined by Einstein's  equations.

The  confinement hypothesis  as  applied to  all perturbed geometries (and not just to the background) can  be  implemented simply as 
\be
\alpha_{*}T^{*}_{\mu\nu} =-8\pi  G T_{\mu\nu},\;\; T^{*}_{\mu a }=0,\;\;T^{*}_{a b}=0  \label{eq:simpleconfinement}
\ee
where $T_{\mu\nu}$  denotes  the  energy-momentum  tensor of 
ordinary matter and  gauge fields, which remain  confined  to  the brane-world. As  we  should expect,  \rf{BE1} reproduces the ordinary Einstein's  equations  when   the extrinsic geometry components are  neglected.

\phantom{x}\hrulefill

\begin{center}
{\textbf{Appendix  B:\\ The  $Z_2$ Symmetry and the Israel-Darmois-Lanczos Condition}}
\end{center}

Here  we use   essentially the same  procedure  as in  Israel's  paper  \cite{Israel},  adapted to  the  case  of  a brane-world  in a constant  curvature bulk.
 The  starting point is 
  Einstein's equation for the bulk geometry \rf{bulkEE2}, now  written as 
 \begin{equation}
{\mathcal{R}}_{AB}=\alpha_{*}(T_{AB}^{*}-\frac{1}{2+N}T^{*}g_{AB})
\label{eq:bulkEE3}
\end{equation}
 For   $D=  4+ N=5$,  the    bulk metric  written in the  embedding vielbein  is  (just for  clarity here  we  set  
    $g_{55}=\varepsilon =1$)
 \[
{\cal G}_{AB}=\left( 
\matrix{g_{\mu\nu}  &   0 \cr
0  &   1} \right)
\]
After explicitly writing  the vielbein components of the Ricci tensor
${\mathcal{R}}_{AB}$  in the case of  the constant  curvature bulk,  we  find from  \rf{BE1} that 
 \begin{eqnarray}
&&\phantom{x}\hspace{-10mm}\!\! R_{\mu\nu}\!  -\!\!\frac{\partial
k_{\mu\nu}}{\partial y}\!-\!\! 2k_{\mu}^{\rho}k_{\rho\nu} \!\! +\!\!h
k_{\mu\nu}\!\! =\!\!\alpha_{*}(T^*_{\mu\nu}\!\!-\!\!\frac{1}{3}T^* g_{\mu\nu})
 \label{eq:R+}
 \end{eqnarray}
Now,  consider  that  the background   $y=0$  separates  two   sides  of the bulk, labeled   by   $+$ and $-$  respectively,
  and find the value of   \rf{R+} as we approach  $y=0$  from each side.
   
  Like in  \cite{Israel}, we  consider 
   two  distinct  situations: Case  (i)  is  characterized by a  continuity  of the extrinsic   curvature  across the  boundary $y=0$: $k_{\mu\nu}^+ =  k_{\mu\nu}^-$,    with the supposition that  the  confined matter    is 
   given by a well defined differentiable  energy-momentum tensor.  Nothing else is  added. Then,  in the  constant curvature bulk the
   equations equivalent to the O'Brien-Synge junction conditions   \rf{BE2}  and  \rf{BE3},    are  just  identities. As  for the tensor equation  \rf{BE1}, we notice that    the value  of $ R_{\mu\nu}$  is  the same  on both  sides of $\bar{V}_4$.  On the other  hand, admitting that the brane-world is orientable, then
    the  term $(-\!\frac{\partial
k_{\mu\nu}}{\partial y})$ and all  terms  involving  the square of  $k_{\mu\nu}$ do not change sign across the boundary.  Since in this case the  confined matter  is intrinsic and well defined, it follows that 
the  differences of  \rf{R+} calculated on both sides of the boundary   cancel each other as  $y \rightarrow 0$.

This  situation changes in case (ii),     characterized by  a  jump  in the extrinsic  curvature  $k_{\mu\nu}^+ \neq  k_{\mu\nu}^-$   across a  background  caused by  a confined distributional  source.
In this case, the  derivatives $(\!\frac{\partial k_{\mu\nu}}{\partial y}) $  in   \rf{R+}  continuously changes  as it approaches  $y=0$, 
so  that 
the  difference between the  values  of \rf{R+} calculated from  one to the other side  of the background is
 \begin{equation}
 -[(\frac{\partial
k_{\mu\nu}}{\partial y})]
=\alpha_{*}\left([T]_{\mu\nu}-\frac{1}{3}[T]g_{\mu\nu} \right)
\label{eq:mirror}
\end{equation}
where we have denoted   $[X]=(X^{+}-X^{-})$.
Since we cannot  anticipate  the  value of  the left hand  side of  \rf{mirror} as   $y \rightarrow 0$,   we may  apply the mean value theorem  for  the differentiable tensor  $k_{\mu\nu}$ in  the interval   $[-y,y]$, to  obtain
 $-[(\!\frac{\partial
k_{\mu\nu}}{\partial y})]
= \frac{-{k}^+_{\mu\nu}  +{k}^-_{\mu\nu}}{y}$. 
To evaluate  the differences of  the  right hand  sides of    \rf{mirror},   we  may express   $T_{\mu\nu}$  as  a delta  function,   noting that for
  $X=\bar{X}(x)\delta(y)$  we have
\begin{eqnarray*}
y[X]= \int_{-y}^{y} \frac{d}{d\xi}\left(|\xi|{X}\right) d\xi
=\int_{-y}^{y}|\xi| \frac{dX}{d\xi} d\xi  + \int_{-y}^{y} \frac{\partial |\xi|}{\partial \xi}Xd\xi \\
=\int_{-y}^{y} |\xi|\bar{X}\delta'(\xi) d\xi  + \int_{-y}^{y}\bar{X}\frac{\partial |\xi|}{\partial\xi}\delta(\xi)d\xi =2 \bar{X}
\end{eqnarray*}
In particular,   for
$\bar{X}=(\bar{T}_{\mu\nu}-1/3\bar{T} \bar{g}_{\mu\nu})$,   we obtain  Lanczos'  equation  describing the jump
   of the extrinsic curvature
\begin{equation}
{k}_{\mu\nu}^+ -{k}_{\mu\nu}^- =-2\alpha_{*}(\bar{T}_{\mu\nu}-\frac{1}{3}\bar{T}\bar{g}_{\mu\nu})
\label{eq:Lanczos}
\end{equation}
To  obtain the IDL  condition  we need to specify how  $k_{\mu\nu}$ changes from one  side to the other. This  is  precisely what the   $Z_2$  symmetry does, where
 the background  $y=0$  acts  as  a mirror  for all objects  that  sense the  extra  dimension. The normal   $\eta_{,\mu}$  and its   derivatives    have  inverted  mirror images,  so  that   from the   definition \rf{kmunua}, the jump 
 of the  extrinsic  curvature  is 
\be
k^{+}_{\mu\nu}=-k^{-}_{\mu\nu} \label{eq:jump}
\ee
so that   equation   \rf{Lanczos} gives at  $y=0$  the  Israel-Darmois-Lanczos  condition
\begin{equation}
\bar{k}_{\mu\nu} =-\alpha_{*}(\bar{T}_{\mu\nu}-\frac{1}{3}\bar{T}\bar{g}_{\mu\nu})
\label{eq:israel}
\end{equation}
specifying the  value  of the  extrinsic  curvature of  the background in terms of the   energy-momentum tensor  of its confined sources. 
 
  Reciprocally,  using the  definition
of $k_{\mu\nu}$,  the  values  calculated on  the  two  sides of the background  are   $k_{\mu\nu}^{\pm}=  -{\cal  Z}_{,\mu}^{A}\eta^{\pm B }_{\;\; ,\nu}\,{\cal G}_{AB}$,  and  therefore   \rf{jump} implies that   $\eta^{A +}_{,\mu}=-\eta^{A -}_{,\mu} $,  or  in other  words  that  the  background behaves  as  a mirror  for the derivatives  of  $\eta$. We  conclude  that  while  \rf{Lanczos}
   follows from  Einstein's  equation  of the bulk  plus the  distributional  source
   of the brane-world, the   $Z_2$  symmetry  (or  any  symmetry  producing the same mirror effect) completely specifies $\bar{k}_{\mu\nu}$
   in terms of the  confined  source $\bar{T}_{\mu\nu}$.

One  aspect  that has not been  considered 
is that   the  IDL   condition, which  was originally applied to  two or three dimensional  hypersurfaces,  may 
 imply in a limited class of  admissible  background  brane-worlds in a higher-dimensional  bulk, depending  on  how  general is that 
  confined  source.    For example,  if  we  consider  a confined source   such as  $\bar{T}_{\mu\nu}= 1/3\bar{T}\bar{g}_{\mu\nu}$, then from  \rf{israel} it follows that $\bar{k}_{\mu\nu}=0$
which means  that the background  is  just a  plane. But  from  \rf{kmunua} it
follows that  all   perturbations also 
 have  zero  extrinsic  curvatures  and  consequently they are  also   planes. 
 Some other examples  are  discussed  in     \cite{Szekeres,Collinson}

Another mathematical aspect  which  deserves further attention 
 is the fact  that  with  the  $Z_2$ symmetry, 
 for each perturbation of  the background  there will be a mirror  image on the other  side of it.
Therefore,   to  each point of the background  corresponds  two    different    tangent  vectors, one on  each side,  whose projections on the  background  give  vectors  pointing  on opposite directions.  This means that  the  derivatives  of the embedding  functions are  not well defined.  Under this  condition 
 the perturbations  of the background 
 cannot  be guaranteed to  remain an  embedded  differentiable manifold \cite{Nash}.


\begin{thebibliography}{20}

\bibitem{Perlmutter}  
S. Perlmutter  et al., \apj, {\bf{517}}, 565, (1999); A. Rises   et all, Astron. J. {\bf{116}}, 1009 (1998).

\bibitem{Weinberg}  
S. Weinberg, Rev. Mod. Phys. {\bf 61}, 1, (1989);  S.
Weinberg, astro-ph/9610044.

\bibitem{Straumann1} 
N. Strumann, proc.  Heidelberg 2000,  110-124, (2000),  astro-ph/0009386.

\bibitem{Caldwell} 
R. Caldwell, et al,  Phys. Rev. Let.  {\bf{80}}, 1582 (1998).

\bibitem{Carroll}  
S. M. Carroll,  astro-ph/0107571.

\bibitem{XCDM} 
M. S. Turner and M. White, Phys. Rev. {\bf{D56}}, R4439
(1997); T. Chiba, N. Sugiyama  and T. Nakamura,  Mon. Not. Roy.
Astron. Soc., {\bf{289}}, L5 (1997); J. S. Alcaniz and J. A. S.
Lima, \apj, {\bf{550}}, L133 (2001) (astro-ph/0109047); Z.-H. Zhu,
M.-K. Fujimoto and D. Tatum, Astron. Atrophys. {\bf{372}}, 377
(2001); A. R. Cooray  \& D. Huterer, Astrophys. J.  \textbf{513},  l95 (1999);
P. S.  Corasaniti et al, Phys. Rev.  \textbf{D70}, 083006  (2004);  Y.Wang  \&  M. Tegmark, Phys. Rev. Lett.  \textbf{92}, 241302-1 (2004) 


\bibitem{ADD}
N. Arkani-Hamed et Al,  Phys. Lett. {\bf{B429}}, 263
(1998), Phys. Rev. Lett. {84}, 586, (2000).

\bibitem{RS} 
L. Randall \& R. Sundrum,  Phys. Rev. Lett. {\bf{83}}, 3370,(1999);
L. Randall and R. Sundrum,  Phys. Rev. Lett.  {\bf 83}, 4690
(1999).

\bibitem{Royreview} 
R.  Maartens, Living Rev.Rel.7:1-99,2004, gr-qc/0312059.
\bibitem{LangloisReview}  D. Langlois,
 gr-qc/0410129
\bibitem{Olinto} 
 Eun-Joo Ahnn, Maximo Ave, Marco Cavaglia, Angela V. Olinto,  Phys. Rev. \underline{D68}, 043004,(2003),  hep-ph/0306008,  also, 
    Phys.Lett. \underline{B551},1, (2003),
 hep-th/0201042

\bibitem{DimopoulosLandsberg}  S. Dimopoulos \& G. Landsberg,   hep-ph/0106295

\bibitem{Cheung} K.  Cheung, DPF annual meeting,  APS,  (2003),    hep-ph/0305003

\bibitem{CavagliaRoy} 
M. Cavaglia, S. Das  and R. Maartens, Class. Quant. Grav. {\bf{20}},
L205 (2003).


\bibitem{Lorenzana} 
A. P\'erez-Lorenzana,  $9^{th}$ Mexican School on Particle and  Fields, Metepec  2000, 53-85  (2000), hep-ph/0008333.




\bibitem{Binetruy1} 
P.  Binetruy, C. Deffayet and D. Langlois, Nucl. Phys. {\bf{B565}},
269, (2000). hep-th/9905012.
\bibitem{Binetruy2} 
P. Binetruy et al, Phys. Lett.  {\bf B477}, 285 (2000),  hep-th/9910219.

\bibitem{Cline}  
J. M.  Cline  et  al,  Phys.  Rev. Lett. {\bf{83}}, 4245 (1999).


\bibitem{TsijukawaLiddle} 
S. Tsujikawa \& A. R. Liddle,  ICAP, 0403, 001, (2004), astro-ph/0312162.
\bibitem{MaiaCR}  
M. D. Maia,  astro-ph/0404370

\bibitem{TsijukawaRoy}  S. Tsujikawa, M. Sami \& R. Maartens, Phys. Rev. \textbf{D70}, 063525, (2004), astro-ph/0406078

\bibitem{Bratt}  J. D.  Bratt  et al, Phys. Lett.{\bf  D70},  063525, (2004)
\bibitem{RingevalDurrer} 
C. Ringeval, T. Boehm, R. Durrer hep-th/0307100.


\bibitem{DGP} 
G. Dvali, G. Gabadadze \&  M. Porrati, Phys. Lett. {\bf{B485}}, 208 (2000)hep-th/0005016.
\bibitem{DvaliTurner} G. Dvali  \&  M. S. Turner, astro-ph/0301510
\bibitem{MaiaFR} M. D. Maia  et al,  Int.J.Mod.Phys {\bf A17}, 4355, (2002),
gr-qc/0212059 



\bibitem{Davis} Anne-Christine Davis  et all, Phys.  lett.  \textbf{B504}, 254, (2001),  hep-ph/0008132
\bibitem{Gergely}  L.A.Gergely,  
Phys. Rev. \textbf{ D68}, 124011  (2003),   gr-qc/0308072

\bibitem{Bowcock}P. Bowcock,  C. Charmosis \&  R. Gregory, Class. Quant. Grav.  \textbf{17},  4745, (2000),
hep-th/0007177

\bibitem{CarterUzan}  B. Carter  \&  J. P.  Uzan, Nucl. Phys. \textbf{ B606}, 45, (2001), gr-qc/0101010

\bibitem{Deruelle} 
N. Deruelle, T. Dolezel and J. Katz, Phys. Rev. {\bf{D63}}, 083513
(2001) hep-th/0010215.

\bibitem{BattyeCarter} R. A. Battye \&  B. Carter,
 Phys. Lett. {\bf{B509}}, 331, (2001),
  hep-th/0101061.

\bibitem{BattyeMennim}  R. A. Battye et al, Phys.  Rev.  {\bf D64}, 124007,  (2001).



\bibitem{MaiaGB} 
M. D.  Maia  \&  E. M.  Monte,  Phys. Lett. {\bf{A297}}, 9 (2002).

\bibitem{MaiaAC} 
M. D. Maia, E. M. Monte, J. M. F. Maia , Phys.Lett. { \bf B585}, 11, (2004), astro-ph/0208223.
\bibitem{Nash} 
J. Nash, Ann.  Maths. {\bf{63}}, 20, (1956).

\bibitem{Greene} 
R. Greene, Mem. Amer. Math. Soc.   \textbf{97} (1970).

\bibitem{Maeda}  
K. Maeda, Phys. Rev. {\bf{D64}}, 123525 (2001). astro-ph/0012313.

\bibitem{Rosen}  
J. Rosen. Rev. Mod. Phys. {\bf {37}, 204, (1965)}.
\bibitem{MaiaMI} 
M. D. Maia \& W. L. Roque Phys. Lett. {\bf {A139}}. 121, (1989).


\bibitem{Sahni}  V. Sahni \& Y. Shtanov,
JCAP, \textbf{ 0311}, 014, (2003), 
astro-ph/0202346,



\bibitem{triess} 
M. S. Turner \& A. G. Riess, \apj {\bf{569}}, 18 (2002).



\bibitem{riess} 
M. S. Turner \& A. G. Riess, \apj {\bf{607}}, 665 (2004).

\bibitem{garn}  
P. M. Garnavich {\it et al.}, \apj {\bf{509}}, 74 (1998).

\bibitem{wang} 
L. Wang  et al \apj {\bf{530}}, 17 (2000).

\bibitem{chae} 
 Kyu-Hyun Chae {\it et al.}, Phys. Rev. Lett. {\bf{89}},
151301 (2002).

\bibitem{abha} 
A. Dev, D. Jain \& S. Mahajan, Intl. J. Mod. Phys. \textbf{ D13}, 1005, (2004), astro-ph/0307441.
\bibitem{schu} 
P. Schuecker et al, Astron-Astrophys.\textbf{ 402}, 53 (2003), astro-ph/0211480.
\bibitem{alcaniz2003} 
J. S. Alcaniz, Phys. Rev. \textbf{D69}, 083521 (2004), astro-ph/0312424.
\bibitem{alcaniz2003b} J. S. Alcaniz, Phys. Rev. {\bf{D69}}, 083521 (2004)




\bibitem{jsa} 
J. A. S. Lima \& J. S. Alcaniz, \apj {\bf{566}}, 15 (2002).

\bibitem{cora} 
P. S. Corasaniti \& E. J. Copeland, Phys. Rev. {\bf{D65}},
043004 (2002).

\bibitem{Gupta} 
S. N.  Gupta, Phys. Rev.  {\bf 96}, 1683, (1954).


\bibitem{Pavsic}
M. Pavsic and V. Tapia, gr-qc/0010045.

\bibitem{Eisenhart}  
L. P.  Eisenhart,  {\em Riemannian Geometry}, Princeton  U.P. Reprint  (1966).
\bibitem{Israel}   W. Israel,  Il  Nuovo Cimento,  \textbf{44}, 4349, (1966).
\bibitem{Szekeres} P. Szekeres,  Il Nuovo Cimento,  {\bf 43}, 1062  (1966)
\bibitem{Collinson} C. D. Collinson,  J. Phys. A  {\bf 4}, 206 (1971)


\end{thebibliography}
\end{document}